\documentstyle[12pt,epsfig]{article}

\def\gsim{\lower0.5ex\hbox{$\:\buildrel >\over\sim\:$}}
\def\lsim{\lower0.5ex\hbox{$\:\buildrel <\over\sim\:$}}

\def\ocal{{\cal O}}

\def\lesim{\,{\raise-3pt\hbox{$\sim$}}\!\!\!\!\!{\raise2pt\hbox{$<$}}\,}

\def\as#1{\alpha_{{\rm s}#1}}
\def\av{\alpha_{\rm v}}
\def\awl{\alpha_{{\rm w}l}}
\def\awr{\alpha_{{\rm w}r}}

\def\mw{m_{\rm w}}
\def\gw{\Gamma_{\rm w}}
\def\vevof#1{\left\langle#1\right\rangle}

\begin{document}
\baselineskip=18pt

\begin{flushright}
{UG-FT-230/08}\\
{CAFPE-100/08}\\
{UCI-TR-2008-21}\\
{BNL-HET-08/14}\\
{UCRHEP-T451}\\
\end{flushright}

\begin{center}
{\large\bf
Heavy Majorana Neutrinos in the
Effective Lagrangian Description: Application to Hadron Colliders}
\end{center}

\vspace*{0.5 in}

\renewcommand{\thefootnote}{\fnsymbol{footnote}}
\centerline{Francisco del \'Aguila$^a$\footnote{Electronic address: faguila@ugr.es},
Shaouly Bar-Shalom$^{b,c}$\footnote{Electronic address: shaouly@physics.technion.ac.il},
Amarjit Soni$^d$\footnote{Electronic address: soni@bnl.gov},
Jose Wudka$^{a,e}$\footnote{Electronic address: jose.wudka@ucr.edu}}

\vspace{0.1 in}

\centerline{\it $^a$Departamento de F{\'\i}sica Te\'orica y del Cosmos and CAFPE,}
\centerline{\it Universidad de Granada, E-18071 Granada, Spain}
\centerline{\it $^b$Physics Department, Technion-Institute of Technology, Haifa 32000, Israel}
\centerline{\it $^c$Department of Physics and Astronomy, University of California, Irvine, CA 92697, USA}
\centerline{\it $^d$Theory Group, Brookhaven National Laboratory, Upton, NY 11973, USA}
\centerline{\it $^e$Department of Physics, University of California, Riverside, CA 92521, USA}

\renewcommand{\thefootnote}{\arabic{footnote}}
\setcounter{footnote}{0}
\vspace*{0.5 in}
\begin{abstract}
We consider the effects of heavy Majorana neutrinos $N$ with sub-TeV
masses.  We argue that the mere presence of these particles would be a
signal of physics beyond the minimal seesaw mechanism and their
interactions are, therefore, best described using an effective Lagrangian. We then
consider the complete set of leading effective operators (up to dimension 6)
involving the $N$ and
Standard Model fields and show that these interactions can be
relatively easy to track at high-energy colliders.
For example, we find that an exchange of a TeV-scale heavy vector field
can yield thousands of characteristic same-sign lepton number violating $\ell^+ \ell^+ jj$ events
($j=$ light jet) at the LHC if $m_N \lsim 600$ GeV, which can also have a distinctive
forward-backward asymmetry signal; even the Tevatron has good prospects for this signature if
$m_N \lsim 300$ GeV.
\end{abstract}

\vspace*{0.5 in}

The spectacular discovery in the past decade of neutrino oscillation
and its interpretation in terms of a non-vanishing neutrino mass matrix is one
of the most important recent discoveries in particle physics. The
$m_\nu \gsim O(10^{-2})$ eV neutrino masses that appear in this scenario
are difficult to generate naturally in the
 Standard Model (SM)
using the Yukawa interactions; the sub-eV mass scale then suggests the
presence of new physics (NP) beyond the SM.
One attractive framework for generating light neutrino masses
naturally is the so-called seesaw mechanism, which requires
the presence of one or more heavy right-handed neutrino species
$N_a$ with interactions of the form
\begin{eqnarray}
{\cal L}_{\nu SM} \equiv {\cal L}_{SM} + \left( \frac{1}{2}
\bar N_a M_{a b} N_b^c - \bar L_i \tilde\phi
Y_{i a} N_a + \hbox{H.c.} \right) ,
\label{nSMF}
\end{eqnarray}
where $L$ denotes the left-handed SU(2) lepton doublet, $\phi$ the
SM scalar isodoublet, $Y$ the Yukawa coupling matrix,
and $M$ the Majorana mass matrix.
If the structure of
$M$ does not allow for a conserved fermion number
then the heavy neutrinos are of Majorana type
and they exhibit characteristic lepton number
violation (LNV) effects that have very distinctive observable
signatures.

In this case, the light neutrino mass matrix is
\begin{eqnarray}
m_\nu = - m_D M^{-1} m_D^T, \quad m_D = Y \vevof\phi  = Y \frac{v}{\sqrt{2}}
\label{seesawmass}~.
\end{eqnarray}
so that $m_\nu\sim 0.01 $ eV if, for example,
$M \sim 100$ GeV and $ m_D \sim m_{\rm electron}/10$
or if $M \sim 10^{15} $ GeV and $m_D \sim m_W$.
The second choice, which seems to be favored by naturalness (since
then $Y \sim O(1)$), clearly leads to the decoupling of the $N$.  In fact, even
if $M \sim 100$ GeV, such that $Y$ is
fine-tuned to the level $\sim 10^{-7}$, we expect $N$ to
decouple since (\ref{seesawmass}) necessarily leads to a vanishingly
small $N-\nu_L$ mixing, $U_{\ell N} \sim \sqrt{m_\nu/M} \sim 10^{-7}
$, and this parameter governs all interactions of $N$ with the SM
particles, e.g., the V$-$A $\ell N W$ vertex \cite{delAguila2}:
\begin{eqnarray}
{\cal L}_{V-A}^W = -\frac{g}{\sqrt{8}} U_{\ell N} \overline{N^c} \gamma^\mu
(1-\gamma_5) \ell W^+_\mu + {\rm H.c.}
\label{YWln} ~.
\end{eqnarray}
Thus, any LNV signal of an EW-scale $N$ would unambiguously indicate the
existence of NP beyond the minimal seesaw framework encoded
in ${\cal L}_{\nu SM}$; the study of heavy Majorana neutrino physics
is then of central importance for our understanding of the short
distance dynamics underlying EW physics.

In this letter we will thus consider $N$ interactions and
phenomenology in the Majorana scenario
when $ M $ is relatively light, $M \lesim O(1) $ TeV, and
its mixing with $\nu _L$ negligible.
Our primary purpose here
is to present a natural, model-independent formalism
that allows a broader and a more reliable
view of the expected physics of heavy TeV-scale Majorana neutrinos,
and lays the ground for further investigations of $N$-mediated LNV
phenomenology at high-energy colliders. We will give a
complete set of leading effective operators (up to dimension 6)
involving the $N$ and
SM fields and then, as an illustration, use it
to demonstrate some aspects of $N$-phenomenology at present or near
future high-energy colliders, such as the Tevatron and the
LHC.
Our approach departs from the traditional viewpoint (see, e.g.
\cite{delAguila2,delAguila1,han,Wpapers,old-N-pheno}), where the
couplings in (\ref{YWln}) (and the associated $\nu N Z$ and $\nu N
H^0$ interactions) was assumed to determine the rate of $N$-mediated
LNV signals and to satisfy $ U_{\ell N} \lesim {\cal
O}(0.1)$, i.e., many orders of magnitude larger than the value
$\sim {\cal O}(10^{-7})$ derived from the seesaw
mechanism (\ref{seesawmass}).
Although there are models that can accommodate this scenario
\cite{beyond}, they usually rely on fine
tuning or on an extended spectrum,
as it is difficult to meet these conditions otherwise.

The effects of the NP underlying ${\cal L}_{\nu SM}$ can be
parameterized by a series of effective operators ${\cal O}_i$
constructed using the $\nu$SM fields and whose coefficients are
suppressed by inverse powers of the NP scale $\Lambda$,
\begin{eqnarray}
{\cal L} = {\cal L}_{\nu SM} + \sum_{n=5}^\infty
 \frac{1}{\Lambda^{n-4}} \sum_i \alpha_i {\cal O}_i^{(n)}
 \label{effLag}~,
\end{eqnarray}
where $n$ is the mass dimension
of $ {\cal O}_i^{(n)}$ (we assume decoupling and
weakly coupled heavy physics,
so $n $ equals the canonical dimension).
Dominating NP effects are generated by contributing operators
with the lowest $n$ value that can be generated at tree level.
The complete list of baryon and lepton number conserving effective operators
of dimension below 6 involving only SM fields are listed in
\cite{effectiveL}; some LNV operators constructed with SM fields are listed in
\cite{effLNV}. Those involving also $N$ are listed below.

There are two tree-level-generated (TLG) dimension $5$ operators
involving the neutrinos, $ (\bar L \tilde \phi)(\phi^\dagger L^c) $
first presented in \cite{Weinberg:1979sa}, and a new one: $ (\bar N
N^c) (\phi^\dagger \phi)$.  Both these terms violate lepton number;
the effects of the second one on the reactions studied below can be
absorbed in a renormalization of the Majorana mass $M$. The dimension
6 TLG operators can be sub-divided into those involving
scalars and vectors (we will use $e,u,d$ and $L,Q$ to denote the
right-handed SU(2) singlets and left-handed SU(2) doublets,
respectively):
\begin{eqnarray}
\ocal_{L N \phi} = (\phi^\dagger \phi) (\bar L N \tilde\phi),~
\ocal_{N N \phi} = i (\phi^\dagger D_\mu \phi) (\bar N \gamma^\mu N),~
\ocal_{N e \phi} = i (\phi^T \varepsilon D_\mu \phi)
(\bar N \gamma^\mu e)~,
\label{FSV}
\end{eqnarray}
and 4-fermion contact terms that either conserve baryon-number
(here $f=u,d,Q,N,e$ or $L$):
\begin{eqnarray}
\ocal_{d u N e} = (\bar d \gamma^\mu u)(\bar N \gamma^\mu e) , &&
\ocal_{f NN} = (\bar f \gamma_\mu f)(\bar N \gamma^\mu N), \cr
\ocal_{LNLe} = (\bar L N) \varepsilon (\bar L e) , &&
\ocal_{LNQd} = (\bar L N) \varepsilon (\bar Q d) , \cr
\ocal_{Q u N L} = (\bar Q u)(\bar N L) , &&
\ocal_{QNLd} = (\bar Q N)\varepsilon (\bar L d) , \cr
\ocal_{L N} = |\bar L N|^2 ,  &&
\ocal_{Q N} = |\bar Q N|^2 , \cr
\ocal_{NN} = (\bar N N^c)^2 , &&
\ocal_{NN}' = |\bar N N^c|^2 \ ,
\label{4FBC}
\end{eqnarray}
or violate baryon-number by one unit:
\begin{eqnarray}
\ocal_{QdN} = (\bar Q Q^c)(\bar d N^c) , &&
\ocal_{QNdQ} = (\bar Q N^c)(\bar d Q^c) , \cr
\ocal_{uNd} = (\bar u N^c)(\bar d d^c) , &&
\ocal_{u d d N} = (\bar u d^c)(\bar d N^c) \ .
\label{4FBV}
\end{eqnarray}
In addition, there are loop-generated operators whose coefficients are
naturally suppressed, $\alpha \sim \ocal (1/16\pi^2)$:
\begin{eqnarray}
\ocal^{(5)}_{NNB} & = & \bar N \sigma^{\mu\nu} N^c B_{\mu\nu}, \cr
\ocal_{ N B} = (\bar L \sigma^{\mu\nu} N) \tilde \phi B_{\mu\nu} , &&
\ocal_{ N W } = (\bar L \sigma^{\mu\nu} \tau^I N) \tilde \phi W_{\mu\nu}^I , \cr
\ocal_{ D N} = (\bar L D_\mu N) D^\mu \tilde \phi, &&
\ocal_{ \bar D N} = (D_\mu \bar L N) D^\mu \tilde \phi \ .
\label{loopO}
\end{eqnarray}

The above operators can give rise to a rich $N$-collider phenomenology.
In this paper we will focus only on $N$-signals at hadron colliders.
Specifically, we will consider the widely studied Drell-Yan like
production of the $N$ in association with a charged lepton: $ p
\bar p , pp \to N \ell $, followed by the decays $N \to \ell jj$ ($j$
stands for a light-quark jet), which gives a distinct LNV signal:
same-sign charged leptons in association with a pair of light
jets\footnote{We focus on the positively-charged
di-lepton signal; at the Tevatron $\sigma(p \bar p \to
\ell^+ \ell^+ jj) = \sigma(p \bar p \to \ell^- \ell^- jj)$ while at
the LHC $\sigma(p p \to \ell^+ \ell^+ jj) \sim 2\ \sigma(p p \to \ell^-
\ell^- jj)$.}
\begin{eqnarray}
p \bar p , pp \to \ell^+ \ell^+ jj \ ,
\label{process}
\end{eqnarray}
which is traditionally taken to be the leading
$N$-signature at the LHC \cite{delAguila2,delAguila1,han,Wpapers},
since it is expected to be the easiest to detect.  However, all
previous studies on this signal assumed that the underlying hard
process is $u \bar d \to W^{+ \star} \to N \ell_L^+$ followed by $N
\to W^- \ell_L^+ \to jj \ell^+_L$, with an unnaturally large coupling
$U_{\ell N} \sim {\cal O}(0.1)$ in (\ref{YWln}).
In contrast, we will see that the
effective Lagrangian description outlined above suggests that the
$\ell^+ \ell^+ jj$ signature is expected to be dominated by other operators.

The TLG operators $ {\cal O}_i $ that contribute to the process (\ref{process})
correspond to $ i = N e\phi$, $duNe$, $QuNL$, $LNQd$ and $QNLd$,
so that (after spontaneous symmetry breaking) the relevant terms in
the effective theory are
\begin{eqnarray}
{\cal L}_{eff}^{N} &=& \frac{1}{\Lambda^{2}} \left[ - \sqrt{2} v \mw
  \left(\awl \overline{N^c} P_R + \awr \bar N P_L \right) \gamma^\mu e W^+_\mu
  + \av\left(\bar d \gamma^\mu P_R u \right)
  \left( \bar N \gamma_\mu P_R e \right)
  \right. \cr
&& \left. + \ \as1 \left(\bar u P_L d \right) \left( \bar e P_R N
\right) - \as2 \left(\bar u P_R d \right) \left( \bar e P_R N \right)
+ \as3 \left(\bar u P_R N \right) \left( \bar e P_R d \right) + {\rm
H.c.} \right] \ ,
\label{NRinter}
\end{eqnarray}
where $\awr \equiv \alpha_{N e \phi}/2$, $\av \equiv \alpha_{duNe}$,
$\as1 \equiv \alpha_{QuNL}$, $\as2 \equiv \alpha_{LNQd}$ and $ \as3
\equiv \alpha_{QNLd}$. Although
not explicitly indicated, (\ref{NRinter}) will in general
contain non-diagonal flavor interactions that may involve heavy
quarks; we will return to these issues in a future publication.

For comparison with the literature we also
included a general SM-like V$-$A term [see (\ref{YWln})], $U_{\ell N}
\equiv \awl \times v^2 /\Lambda^2$, even though such a coupling is
expected to be $ \sim 10^{-7} $, in which case the corresponding vertex
will have no observable effects.
Thus, the observation of LNV effects associated with
an $N$ with $M = O(100)$ GeV
will be most likely associated with
$ \alpha_i \not= 0 $ for some $i \not= {{\rm w}l} $,
indicative of physics beyond the classic seesaw mechanism.

Underlying the use of the
effective interactions (\ref{NRinter}) is the presumption that this
NP is not directly observable. Nonetheless one can use
observables contributing to (\ref{process})  to extract (or constrain)
the values of the various coefficients in (\ref{NRinter})
and use this information to
restrict the possible types of NP responsible for these
effects.
While a detailed study in this direction lies beyond the scope of
this paper,
we will comment on how this can be done and
on the precision to which these
coefficients can be measured in the LNV reaction
(\ref{process}).

Using (\ref{NRinter}), we find that the differential cross-section for
the hard process $ u \bar d \to N \ell^+$ and the spin-averaged
differential decay width for $N \to \ell^+ j j$ are
respectively
\begin{eqnarray}
\frac{d \hat\sigma}{d c_\theta} &=& \frac{(\hat s -M^2)^2}{128 \pi \; \hat s
\Lambda^4 } \left\{ \as1^2 + \as2^2
- \as2 \as3 (1 + c_\theta) +
\as3^2\Upsilon_+ + 4 \av^2 \Upsilon_- + 16 \left(\awl^2 \Upsilon_-
+ \awr^2 \Upsilon_+ \right) \Pi_{\rm w}(\hat s) \right\} \ , \label{cs} \\
\frac{d\Gamma}{dx} &=& \frac{M}{128 {\pi}^3} \left( \frac M\Lambda
\right)^4 \left\{ \left( \as1^2 + \as2^2 - \as2 \as3 \right) f_{\rm s}
+ \left[ \as3^2 + 4\av^2 + 16 \left( \awl^2 + \awr^2 \right)
\Pi_{\rm w}(2 z M^2) \right] f_{\rm v}\right\} \ ,
\label{width}
\end{eqnarray}
where we assume real coefficients and
\begin{eqnarray}
\Upsilon_\pm = \frac{1}{4}\left[(1 \pm c_\theta)^2 + M^2 \frac{s_\theta^2}{\hat s}\right]~,~
\Pi_{\rm w}(\hat s) \equiv \frac{\mw^4}{(\hat s-\mw^2)^2 + (\mw \gw)^2} ~,
\end{eqnarray}

\noindent $\gw$ denotes the total $W$ width, $c_\theta$
is the cosine of the center of mass
(CM) scattering angle between $\ell$ and $u$,
$\hat s$ the CM energy squared of the hard
process, $ f_{\rm s} (x) = 12 x^2 z$, $f_{\rm v}(x) = 2 x^2(x+3z)$,
with $z=1/2-x$, and $Mx$ is the energy of the $N$ decay lepton
in the $N$ rest frame.
\begin{figure}[htb]
\begin{center}
\epsfig{file=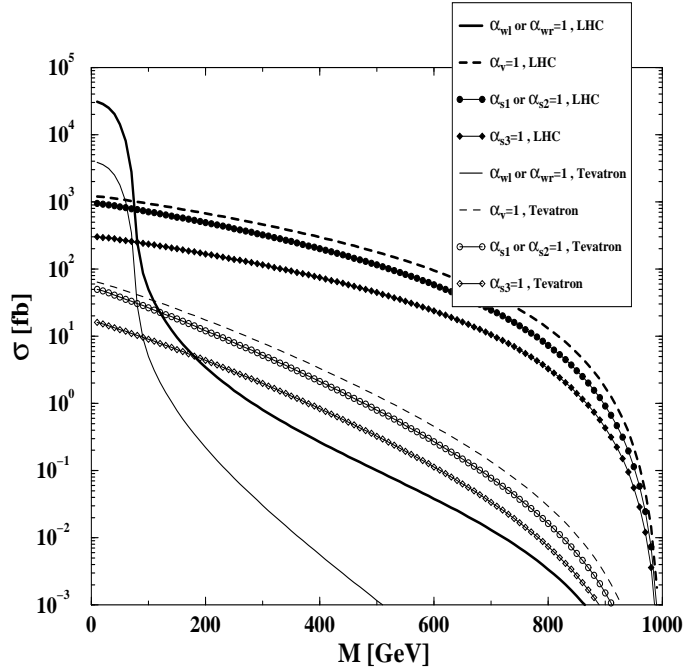,height=9cm,width=9cm,angle=0}
\caption{\emph{The total cross-section $\sigma(p \bar p, pp \to \ell^+
N)$ derived from (\ref{NRinter}) for the Tevatron and the LHC, as a
function of $M$, for $ |c_\theta|<0.9, ~ \Lambda=1$ TeV and $\hat s
< \Lambda^2 $ (for this $ \Lambda $, $ \awl =1 $ corresponds to
$ U_{\ell N} \sim 0.06$). See text.}}
\label{fig1}
\end{center}
\end{figure}

In Fig.~\ref{fig1} we plot the total cross-sections $\sigma$,
convoluted with the initial parton densities inside the (anti)protons, as a
function of $M$ and for $\Lambda=1$ TeV and various values of the
coefficients $\alpha_i$. The cross-section is integrated for
$ |c_\theta| \leq 0.9$ up to $\sqrt{\hat s} < \Lambda $
(the decrease in $\sigma$ with $M$ results from this cut),
imposed in order to insure the validity of the effective
Lagrangian approach.
The signal to background analysis described in
\cite{delAguila2,delAguila1,han,Wpapers} for $pp \to \mu N \to \mu \mu
jj$ also applies to the cross-sections in Fig.~\ref{fig1},
based on which we expect a $5 \sigma$ same-sign leptons
signal at the LHC if $M \lsim 200$ GeV, $\Lambda \sim {\cal O}(1)$ TeV,
and $\awr \sim {\cal O}(1)$ with $ \alpha_i =0 $ otherwise ~\cite{delAguila1}.
The Tevatron is, however,
not sensitive to this process and coupling for $M > \mw$ \cite{delAguila1}.

Also note that the 4-fermion terms can significantly
contribute to $\sigma$, especially for $M > \mw$
when the s-channel W-exchange process is non-resonant.
Hence, if $\av \sim 1$ and $\Lambda \sim 1$
TeV, then $\sigma \gsim 10~(100)$ fb at the Tevatron (LHC)
for $M \lsim 300 ~(600)$ GeV.  Based on
the results of \cite{delAguila1}, such a large $\mu^+ N$
production rate
is within the sensitivity of both colliders
with integrated luminosities of ${\cal O}(10)$
fb$^{-1}$.
The 4-fermion interactions are generated,
e.g., by a new right-handed gauge
interaction mediated by a $W_R^\prime$ too heavy to be directly
observable.\footnote{If the $W_R^\prime$ were to be directly
observable, the sensitivity to $N$ would be markedly
improved \cite{delAguila2},
though the effective theory approach would no longer
be applicable.}

As mentioned previously, one
can use observables such as (\ref{cs}) and (\ref{width}) to
measure or bound the magnitudes of the $ \alpha_i $ in
(\ref{NRinter})
and the $\as2-\as3$ relative phase (terms containing other relative
phases are multiplied by a light lepton or quark mass).
Perhaps the simplest example is the
forward-backward (FB) asymmetry ($A_{FB}$) in the underlying process
$p\bar p, pp \to \ell^+ N$, which requires the
proper lepton assignment, usually that of largest transverse momentum,
and can be used to extract the terms linear in $c_\theta $ in (\ref{cs}).
Note that for the LHC the conventional $A_{FB}$ vanishes
(due to the identical colliding beams) but we can use $A_{FB}^y$,
the double asymmetry in $ \theta $ and the rapidity $y$, as described,
e.g., in \cite{aguila3}.
In Table \ref{tab1} we give the expected FB asymmetries at the Tevatron
and the LHC when a single $ \alpha_i $ is not zero and
for $M=200$ GeV (the asymmetries depend very weakly on $M$).
\begin{table}[htb]
\begin{center}
\begin{tabular}{c|c|c|c|c|c}
~ & \multicolumn{5}{c}{\underline{non-zero coefficient}} 
\\[.1cm]
~ & $\awl$ & $\awr$ & $\av$ & $\alpha_{s1, s2}$
& $\as3$ \\
\hline
$A_{FB}$ (Tevatron) & $0.55$ & $-0.55$ & $0.62$ & $0$ & $-0.62$ \\
$A_{FB}^y$ (Tevatron) & $0.11$ & $-0.11$ & $0.12$ & $0$ & $-0.12$ \\
$A_{FB}^y$ (LHC) & $0.35$ & $-0.35$ & $0.40$ & $0$ & $-0.40$ \\
\end{tabular}
\caption{\emph{The expected FB asymmetries $A_{FB}$ at the Tevatron and
$A_{FB}^y$ at the Tevatron and at the LHC (see text),
corresponding to each of the effective operators in (\ref{NRinter})
when $M=200$ GeV.}}
\label{tab1}
\end{center}
\end{table}
\medskip

Other differential distributions for the reaction (\ref{process})
can also be utilized.
For instance, we find that the invariant mass distribution
of the two leptons or of the two jets,
can differentiate between the $W$ and 4-fermion mediated processes.
On the other hand, by taking the moments of (\ref{width})
with appropriate functions of $x$,
the coefficient combinations multiplying
$ f_{\rm s} $ and $ f_{\rm v} $ can be
measured. Additional information can be extracted from other
differential distributions involving the $N$ spin dependence.
A realistic determination of the constraints on the $ \alpha_i $
requires careful consideration of the various backgrounds and
event selection efficiencies; this lies beyond the scope of the
present work but will be detailed in a future publication.
Here we only remark that
the very distinctive characteristics of LNV signatures of
this type should allow for a drastic reduction of the backgrounds
after an optimal
event (distribution) selection, see e.g., \cite{BarShalom:2008fq}.

To summarize, we have argued
that the natural size of the
heavy-to-light $N-\nu_L$ mixing is expected to be ${\cal
O}(\sqrt{m_\nu/M}) \ll 1 $ within the classic seesaw mechanism,
leading to the decoupling of
the heavy Majorana neutrinos
even if their masses are $\sim 100$ GeV -- $1$ TeV, unless additional
interactions are present. Thus, any signal of EW-scale heavy Majorana
neutrinos provides a strong indication of physics beyond the
minimal seesaw mechanism, at a near-by scale.

Adopting this viewpoint, we re-examined heavy Majorana neutrino physics
using an effective Lagrangian approach. We gave a complete set of
the leading effective operators (of dimension $ \le6 $) involving the $N$ and the SM
fields. As an illustration, we studied the effects
of the higher dimensional operators that yield a new (V$+$A) $\ell N W$
interaction and new 4-fermion
$ud\ell N$ contact terms,
on the LNV process $p \bar p, pp \to \ell^+ N$ followed by $N
\to \ell^+ j j$ at the Tevatron and the LHC.

We found that these new effective operators can significantly enhance
the production of N at hadron colliders, potentially leading to hundreds or
even thousands of LNV $\ell^+ \ell^+ jj$ events at the Tevatron and at
the LHC, if the typical scale of the new physics is $ \Lambda \sim 1$ TeV.
We have also found that it is possible, to a certain extent, to discriminate
between the various types of new physics responsible for the effective
interactions, by measuring
differential distributions of the outgoing charged leptons and
jets. For example, if the new physics is manifest only through a new
(V$+$A) $\ell N W$ interaction, then we expect $\sigma(\ell^+_R \ell^+_R
j j) \gg \sigma(\ell^+_L \ell^+_L j j)$, which would manifest in the
FB asymmetry and will thus stand as an unambiguous signal of beyond
the $N-\nu_L$ mixing scheme implied by the classic seesaw mechanism.

\bigskip

{\it \bf Acknowledgments:} The work of FA and JW was supported in part
by MEC (FPA2006-05294, SAB2006-0173) and by Junta de Andaluc{\'\i}a,
and the work of SBS by NSF
Grants No. PHY-0653656 (UCI), PHY-0709742 (UCI) and by the Alfred
P. Sloan Foundation. The work of AS was supported in part by the DOE grant
DE-AC02-98CH10886 (BNL).
JW is also partially  supported by the U.S. DoE under grant
DE-FG03-94ER40837.

\end{document}